\documentclass[a4paper,fleqn,usenatbib]{mnras}

\usepackage{newtxtext,newtxmath}

\usepackage[T1]{fontenc}
\usepackage{ae,aecompl}

\usepackage{graphicx}	
\usepackage{amsmath}	
\usepackage{amssymb}	
\usepackage{multirow}
\usepackage{array}
\usepackage{subfig}

\title[Visualize galaxy feature with ML]{Visualizing the Hidden Features of Galaxy Morphology with Machine Learning}

\author[J. M. Dai et al.]{
Jia-Ming Dai,$^{1,2}$ \thanks{E-mail: daijiamingdl@gmail.com}
Jizhou Tong$^{1}$
\\
$^{1}$ National Space Science Center, Chinese Academy of Sciences, Beijing 100190,China\\
$^{2}$ University of Chinese Academy of Sciences, Beijing 100049, China\\
}

\date{Accepted XXX. Received YYY; in original form ZZZ}

\pubyear{2018}

\begin{document}
\label{firstpage}
\pagerange{\pageref{firstpage}--\pageref{lastpage}}
\maketitle

\begin{abstract}
We train three convolutional neural networks (CNNs) to classify galaxies with Galaxy Zoo 2 dataset and extract the activations from the last fully connected layer or the last average pooling layer of CNNs to study the high-dimensional abstract feature representations of galaxy images. We apply t-Distributed Stochastic Neighbour Embedding (t-SNE), a popular dimensionality reduction technique, to visualize the high-dimensional galaxy feature representations in two-dimensional scatter plots. From the visualization, we try to understand the galaxy images data itself and obtain some highly valuable insights. For instance, the learned galaxy feature representations from networks indicate that the galaxies belonging to the same class tend to group together, i.e. same morphological galaxies are clustered; The  cluster of completely round smooth galaxy and the cluster of in-between smooth galaxy (between completely round and cigar-shaped) are moved closer, compared to other  clusters; The  cluster of cigar-shaped smooth galaxy and the  cluster of edge-on galaxy are intertwined surprisingly; A galaxy mislabelled as spiral galaxy in the original dataset falls in the  cluster of completely round smooth galaxy, and manual inspection also identifies out the outlier as a completely round smooth galaxy. These findings will facilitate the study of galaxy morphology. 
\end{abstract}

\begin{keywords}
methods: data analysis-techniques: image processing-galaxies: general.
\end{keywords}



\section{Introduction}\label{sec:intro}

Deep learning has been increasingly applied to galaxy morphology classification and achieved a series of success. For example, \citet{dieleman2015rotation} first used CNNs to galaxy morphology classification which exploits galaxy images translation and rotation invariance on Galaxy Zoo 2 dataset. \citet{gravet2015catalog} used the Dieleman model to classify high redshift galaxies in the 5 Cosmic Assembly Near-infrared Deep Extragalactic Legacy Survey (CANDELS). \citet{kim2016star} used CNNs to resolve star-galaxy classification. \citet{aniyan2017classifying} used CNNs to classify radio galaxies into FRI, FRII and Bent-tailed radio galaxies. \citet{sanchez2017improving} used CNNs to improve galaxy morphologies for Sloan Digital Sky Survey(SDSS). These works focus on the classification tasks however we argue that these works can go further by digging into the networks to analyse what features can be learned and how the learned features can help in a better understanding of the data itself. It is well known that the reason for CNNs to have been used so popularly and successfully in galaxy morphology classification lies in that CNNs can be fed with raw data directly and can automatically learn the feature representations needed for classification task \citep{bengio2013representation, lecun2015deep}. These learned feature representations are nonlinear mappings of the original image pixel values, although they are still high-dimensional, they can be used not only for classification, but also for other purposes, such as dimensionality reduction and visualization.

Visualization of high-dimensional data is one of the methods of data analysis and can discover a lot of highly valuable feedbacks. In the last few decades a series of dimensionality reduction techniques have been proposed to visualize high-dimensional data due to the large high dimensional data in real world. Dimensionality reduction techniques include linear techniques, like Principal Components Analysis (PCA) \citep{hotelling1933analysis} and classical scaling \citep{torgerson1952multidimensional}, and nonlinear techniques \citep{van2007introduction,van2009dimensionality}, like Isomap \citep{tenenbaum2000global}, Locally linear Embedding (LLE) \citep{roweis2000nonlinear}, Laplacian Eigenmaps \citep{belkin2002laplacian}, Auto-encoder \citep{hinton2006reducing} and t-Distributed Stochastic Neighbor Embedding (t-SNE) \citep{maaten2008visualizing,van2014accelerating}. t-SNE is intensively applied to visualize high-dimensional data in machine learning due to its capability of preserving the local structure of the data and revealing global structure. \citet{maaten2008visualizing} proposed t-SNE method and demonstrated its good performance with other dimensionality reduction techniques on a wide variety of datasets. \citet{van2014accelerating} trained CNNs to extract image feature representation, used t-SNE techniques to visualize high-dimensional data by scatter plots on MNIST \citep{lecun1998mnist} and SVHN \citep{netzer2011reading} datasets, and obtained some valuable insights and qualitative analysis. Recently, \citet{rauber2017visualizing} systematically trained multilayer perceptrons (MLPs) and CNNs to extract activations from the last fully connected layer in three traditional image classification benchmark datasets (MNIST, SVHN and CIFAR-10), then used t-SNE techniques to visualize the learned presentations in two-dimensional scatter plots. Some highly valuable feedbacks were discovered. 

These work inspires the study of the feature representations of galaxy morphology automatically learned from networks. We look deeply into the networks searching for the physical meanings of the high-dimensional abstract feature representations and show how these feature representations help to understand the galaxy images data itself. In this paper we train three CNNs to classify galaxies into five classes. We use activations from the last fully connected layer or the last average pooling layer of CNNs to make high-dimensional galaxy morphological feature representations and use t-SNE techniques to visualize the high-dimensional galaxy morphological feature representations in two-dimensional scatter plots. This is the first time to apply in galaxy morphology.

The outline of the paper is as follows. In Section \ref{sec:classification}, we introduce our three CNNs architectures, dataset selection, image preprocessing, training protocol, classification results and activations extraction. In Section \ref{sec:visual}, we outline t-SNE techniques and present the visualization of the learned feature representations in details. Our conclusions and suggestions for future work are presented in Section \ref{sec:conclusion}.

\section{Classification Framework}\label{sec:classification}

Deep learning models are composed of multiple nonlinear layers to learn data representation automatically for classification, detection and segmentation tasks. Among them, deep convolutional neural networks (CNNs) have become the dominant approach in image classification task \citep{lecun2015deep}. Since 2012, when \citet{krizhevsky2012imagenet} used a CNN to win the first place of the ImageNet Large Scale Visual Recognition Challenge (ILSVRC), CNNs have achieved a series of breakthroughs in image classification. Now, CNNs have been evolved into many versions, such as AlexNet \citep{krizhevsky2012imagenet}, VGG \citep{simonyan2014very}, Inception \citep{szegedy2015going, ioffe2015batch, szegedy2016rethinking, szegedy2017inception}, ResNets \citep{he2016deep, he2016identity}, DenseNet \citep{huang2016densely} and so on. Full details are available in above literatures and \citet{goodfellow2016deep}.

In this section, we first describe dataset selection, data processing, three CNNs architectures and training tips. Next, the classification results are summarized. At last, we extract the representation learned from networks as galaxy morphological features.

\subsection{Dataset}
The galaxy images in this study are drawn from Galaxy Zoo-the Galaxy Challenge\footnote{https://www.kaggle.com/c/galaxy-zoo-the-galaxy-challenge} from Galaxy Zoo 2 (GZ2). In order to select clean samples, we use the rules of selecting clean samples of GZ2 data release \citep{willett2013galaxy}. For example, to select the spiral, cuts are the combination of $f_{features/disk} \geq 0.430$, $f_{edge-on,no} \geq 0.715$ and $f_{spiral,yes} \geq 0.619$. By this means, we classify galaxies into 5 classes, i.e. completely round smooth galaxy, in-between smooth galaxy (between completely round and cigar-shaped), cigar-shaped smooth galaxy, edge-on galaxy and spiral galaxy, which are referred to as 0, 1, 2, 3, 4, respectively. We loosened the thresholds of smooth galaxy from 0.8 to 0.5 and all others are derived from \citet{willett2013galaxy}, where full details of this are available. Table \ref{tab:dataselection} shows the clean samples selection criterion for every class. Dataset reduced to 28790 images after filtering. Each image is of $424\times424\times3 $ pixels in size. We split them into training set and testing set at a ratio of 9:1, 25911 images are used to train models and the remaining 2879 images to evaluate our models.

\begin{table*}
	\centering
	\caption{Clean samples selection in GZ2. The clean galaxy images are selected from GZ2 data release \citep{willett2013galaxy}, in which thresholds determine well-sampled galaxies. Thresholds depend on the number of votes for a classification task considered to be sufficient. As an example, to select the spiral, cuts are the combination of $f_{features/disk} \geq 0.430$ , $f_{edge-on,no} \geq 0.715$, $f_{spiral,yes} \geq 0.619$.}
	\label{tab:dataselection}
	\begin{tabular}{clcll}
		\hline
		Class & Clean sample & Tasks & Selection & $N_{sample}$ \\
		\hline
		0 & Completely round  & T01  & $f_{smooth} \geq 0.469$   & 8434 \\
        & smooth& T07 &$f_{completely ~round} \geq 0.50$ & \\
        \hline
        1 & In between  & T01  & $f_{smooth} \geq 0.469$   & 8069 \\
        & smooth &T07 &$f_{in~ between} \geq 0.50$ & \\
        \hline
        2 & Cigar-shaped  & T01  & $f_{smooth} \geq 0.469$   & 578 \\
        & smooth & T07&$f_{cigar-shaped} \geq 0.50$ & \\
        \hline
        3 & Edge-on  & T01  & $f_{features/disk} \geq 0.430$   & 3903 \\
        & &T02 & $f_{edge-on,yes} \geq 0.602$ & \\
        \hline
        &   & T01  & $f_{features/disk} \geq 0.430$    &  \\
        4& Spiral &T02 &$f_{edge-on,no} \geq 0.715$ & 7806 \\
        & & T04 &$f_{spiral,yes} \geq 0.619$ & \\
        \hline

	\end{tabular}
\end{table*}

\subsection{Preprocessing}
It can be seen that images are composed of large fields of view with the galaxy of interest in the center. So first step, galaxy images are cropped from center to a range scale $ S=[170,240]$ in training set. Almost main information is contained in the center images and many noises like other secondary objects are eliminated. It reduces the dimension of image and it is also a form of data augmentation for training set. Then, images are resized to $ 80 \times 80 \times 3 $ pixels. A random crop is performed to $ 64 \times 64 \times 3 $ pixels. Next, images are randomly rotated with $ 0^{\circ}, 90^{\circ},180^{\circ},270^{\circ} $ due to rotation invariant of galaxy images and randomly horizontally flipped. Brightness, contrast, saturation and hue adjustment are applied to images. These steps are data augmentation to avoid overfitting.  The last step was image whitening. This are the whole preprocessing pipeline in training. After those steps, images ($ 64 \times 64\times 3 $ pixels) will be used as input of networks in training.

At testing time, preprocessing procedure  only includes center cropping (test scale $ Q=\{ 180,200,220,240\} $),  resizing to $ 80 \times 80 \times 3 $ pixels,  center cropping again, and image whitening. Images ($ 64 \times 64 \times 3 $ pixels) will be used as input of networks in testing.

\begin{table*}
	\centering
	\caption{Our three CNNs architectures. First column represents \lq\lq type\rq\rq~ for CNN 1 and CNN 2, fifth column represents \lq\lq layer name\rq\rq~ for CNN 3.}
	\label{tab:networks architectures}
	\begin{tabular}{ccc||cc} 
		\hline
		Type & CNN 1 & CNN 2 & CNN 3& Layer name\\
		\hline
		convolutional & $6 \times 6,32$ & $3 \times 3,64$ & $6 \times 6,64$ & conv 1 \\
		convolutional & - & $3 \times 3,64$ & - & - \\
		pooling & $2 \times 2$, stride 2 & $2 \times 2$, stride 2 &$2 \times 2$, stride 2  & max-pooling \\
        convolutional& $5 \times 5,64$ & $3 \times 3,128$ &  \multirow{3}{*}{$\begin{bmatrix}1\times 1,128\\3 \times 3, 128\\1 \times 1,512\end{bmatrix}\times 2 $} & \multirow{3}{*}{conv 2} \\
        convolutional & - & $3 \times 3,128$ &  &  \\
        pooling& $2 \times 2$, stride 2 & $2 \times 2$, stride 2 &  &  \\
        convolutional & $3 \times 3,128$ & $3 \times 3,256$ & \multirow{3}{*}{$\begin{bmatrix}1\times 1,256\\3 \times 3, 256\\1 \times 1,1024\end{bmatrix}\times 2 $} & \multirow{3}{*}{conv 3} \\
        convolutional & $3 \times 3,128$ & $3 \times 3,256$ &  &  \\
        convolutional & - & $3 \times 3,256$ &  &  \\
        pooling& $2 \times 2$, stride 2 & $2 \times 2$, stride 2 & \multirow{3}{*}{$\begin{bmatrix}1\times 1,512\\3 \times 3, 512\\1 \times 1,2048\end{bmatrix}\times 2 $} & \multirow{3}{*}{conv 4} \\
        convolutional & - &$3 \times 3,512$  &  &  \\
        convolutional & - & $3 \times 3,512$ &  &  \\
        convolutional & - & $3 \times 3,512$ & \multirow{3}{*}{$\begin{bmatrix}1\times 1,1024\\3 \times 3, 1024\\1 \times 1,4096\end{bmatrix}\times 2 $} & \multirow{3}{*}{conv 5} \\
        pooling  & - & $2 \times 2$, stride 2 &  &  \\
        convolutional & - & $3 \times 3,512$ &  &  \\
        convolutional & - & $3 \times 3,512$ &  &  \\
        convolutional & - & $3 \times 3,512$ &  &  \\
        pooling & - & $2 \times 2$, stride 2 & $4 \times 4$ & avg-pooling \\
        fully-connected & 2048 & 4096 &  &  \\
        fully-connected& 2048 & 4096 &  &  \\
        fully-connected & 5 & 5 & 5 & softmax \\
		\hline
	\end{tabular}
\end{table*}

\subsection{CNNs architectures} 

To extract feature representations from the galaxy images, we train three CNNs, as Table \ref{tab:networks architectures} shows:
\begin{enumerate}
  \item CNN 1: CNN 1 is a 7-layers CNN, including 4 convolutional layers and 3 fully connected layers.  It is a slightly modified Dieleman model \citep{dieleman2015rotation}.
  \item CNN 2: CNN 2 has 16 layers totally, 13 convolutional layers and 3 fully connected layers. It is a slightly modified VGG-16 \citep{simonyan2014very}.
  \item CNN 3: CNN 3 is a modified ResNets \citep{he2016deep, he2016identity}, 26 layers totally, where we decrease the depth and widen the channel. It has 4 convolutional groups: conv2, conv3, conv4 and conv5, respectively. We add dropout after $ 3\times 3 ~convolution $ of every residual unit, to prevent coadaptation and overfitting. Downsampling is performed by the last layers in groups conv2, conv3 and conv4 with a stride of 2.
\end{enumerate}

\subsection{Training}

The activation function of all convolutional layers and fully connected layers (except output layer) is Rectified Linear Units (ReLUs) \citep{nair2010rectified}. The networks are trained to minimize cross entropy loss. We use a batch size of 128.  The initial learning rate is set to 0.1, then decreased by a factor of 10 at 30k and 60k iterations.

For CNN 1, we use GradientDescentOptimizer. We stop training after 72k iterations. Weights are initialized by sampling from zero-mean normal distributions (standard deviation 0.01 for convolutional layers and softmax layer, standard deviation 0.001 for fully connected layers). Biases are initialized to small positive values (0.1 for convolutional layers and softmax layer, 0.01 for fully connected layers). Dropout probability value is 0.5 \citep{srivastava2014dropout}.

For CNN 2, we use GradientDescentOptimizer. We stop training after 42k iterations. All weights are initialized by Xavier initializer \citep{glorot2010understanding}. Biases are initialized to 0 for convolutional layers (0.1 for fully connected layers and softmax layer). The weight loss values of the two fully connected layers are 0.0005. Dropout probability value is 0.5.

For CNN 3, we use MomentumOptimizer with Nesterov momentum of 0.9, the weight decay is 0.0001, dropout probability value is 0.8. We adopt batch normalization (BN) \citep{ioffe2015batch} before activation and convolution, following \citet{he2016identity}. The weights are initialized as in \citet{he2015delving}. We stop training after 72k iterations.

Our implementation is based on Python, Pandas, scikit-learn \citep{Pedregosa2012Scikit}, scikit-image \citep{Van2014scikit} and TensorFlow \citep{abadi2016tensorflow}.

\subsection{Classification results}

Table \ref{tab:acc} summaries the results of test accuracy of different methods. Our results are based on the maximum values of 10-times runs of each testing scale. It is obvious all three models of ours achieve excellent performance.

\begin{table}
	\centering
	\caption{Test accuracy of different methods. Our results are based on the maximum values of 10-times runs of each test scale.}
	\label{tab:acc}
	\begin{tabular}{cc} 
		\hline\hline
		Model & Overall Accuracy(\%)\\
		\hline
		CNN 1 & 94.6528 \\
		CNN 2 &  93.6458\\
		CNN 3 &  95.2083\\
		\hline
	\end{tabular}
\end{table}

\subsection{Activations}  

A subset of 1000 images is extracted from training set, each class contains 200 images, which is hence called training-1000. And a random subset of 1000 images is also extracted from testing set, each class contains 200 images (the cigar-shaped is not enough and only has 58 images, in addition 142 image are extracted from training set), hence it is called testing-1000. For CNN 1, we use the activations in the last fully connected layer as 2,048-dimensional feature representations. For CNN 2, we use the activations in the last fully connected layer as 4,096-dimensional feature representations. For CNN 3, we use the activations in the average pooling layers as 4,096-dimensional feature representations, respectively.

\section{Visualizing the learned feature representations}\label{sec:visual}
In this section, we briefly introduce t-Distributed Stochastic Neighbor Embedding (t-SNE) technique, and then use it to visualize the high-dimensional galaxy morphological feature representations learned from CNNs.

\subsection{t-SNE}
t-Distributed Stochastic Neighbor Embedding (t-SNE) is a popular dimensionality reduction algorithm  presented in \citet{maaten2008visualizing,van2014accelerating}. It is a nonlinear dimensionality reduction method which is especially suitable for the visualization of high-dimensional data in a space of two or three dimension, this is so-called a scatter plot. And it is an extremely broadly applicable technique in machine learning due to its capability of preserving the local structure of the data and revealing the global structure.

A dataset $X$ includes $N$ observations $X=\{ x_1,x_2,\cdots,x_N \}$, where are $ D $-dimensional real vector. The goal of t-SNE is to compute a projection  $Y=\{ y_1,y_2,\cdots,y_N \}$ where the neighborhoods from $D$ are preserved, generally, $ y_i \in \mathbb{R}^d $ corresponds to $ x_i \in \mathbb{R}^D$. Typically, $ d=2~and~D \gg d $.

Firstly, compute conditional probability $ p_{j|i} $ that are proportional to the similarity of datapoint $ x_i $ and datapoint $ x_j $.  $ p_{j|i} $ is high when $x_i$ is near to $x_j$ and $ p_{j|i} $ is defined as

\begin{equation}\label{eq:1}
  p_{j|i}=\frac{exp(-||x_i-x_j||^2/2\sigma_i^2)}{\sum_{k\neq i}exp(-||x_i-x_k||^2/2\sigma_i^2)}.
\end{equation}
where $ \sigma_i$ is the variance of the Gaussian that is centered on datapoint $ x_i $.

Then, define the joint probabilities $p_{ij}$ in the high-dimensional space to be the symmetrized conditional probabilities

\begin{equation}\label{eq:2}
  p_{ij}=\frac{p_{j|i}+p_{i|j}}{2N}.
\end{equation}

Next, in the low-dimensional space, define the joint probabilities $ q_{ij} $ which uses a Student t-distribution with one degree of freedom (which is the same as a Cauchy distribution) as the heavy-tailed distribution

\begin{equation}\label{eq:3}
  q_{ij}=\frac{(1+||y_i-y_j||^2)^{-1}}{\sum_{k\neq l}(1+||y_k-y_l||^2)^{-1}}.
\end{equation}

t-SNE aims at minimizing the following cost function $ C $. And the locations of the points $ y_i $ in the map are determined by minimizing the Kullback-Leibler divergence of the distribution $ Q$ from the distribution $ P $, that is

\begin{equation}\label{eq:4}
  C=KL(P|Q)=\sum_{i}\sum_{j}p_{ij}\log\frac{p_{ij}}{q_{ij}}.
\end{equation}

The gradient of the Kullback-Leibler divergence between $ P$ and the Student-t based joint probability distribution $ Q $ is given by
\begin{equation}\label{eq:4}
  \frac{\delta C}{\delta y_i}=4\sum_{j}(p_{ij}-q_{ij})(y_i-y_j)(1+||y_i-y_j||^2)^{-1}.
\end{equation}

For further more details, we refer to \citet{maaten2008visualizing,van2014accelerating}.

\subsection{Visualizing the representations}

\begin{figure*}
\centering
\subfloat[Map of raw samples on the training-1000 subset.]{
\label{fig:train1000}
\begin{minipage}[t]{0.5\textwidth}
\centering
\includegraphics[scale=0.4]{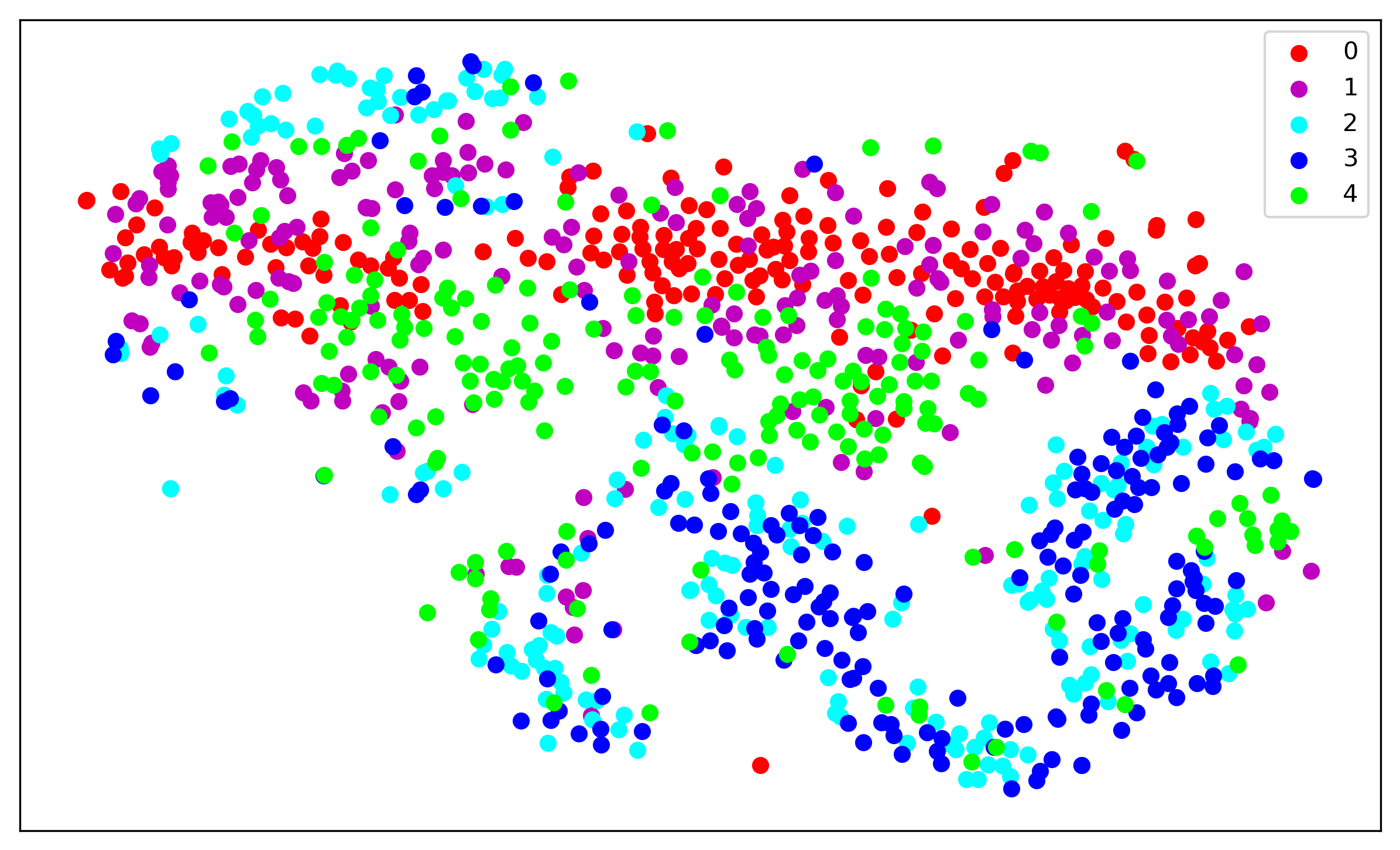}
\end{minipage}
}
\subfloat[Map of raw samples on the testing-1000 subset.]{
\label{fig:test1000}
\begin{minipage}[t]{0.5\textwidth}
\centering
\includegraphics[scale=0.4]{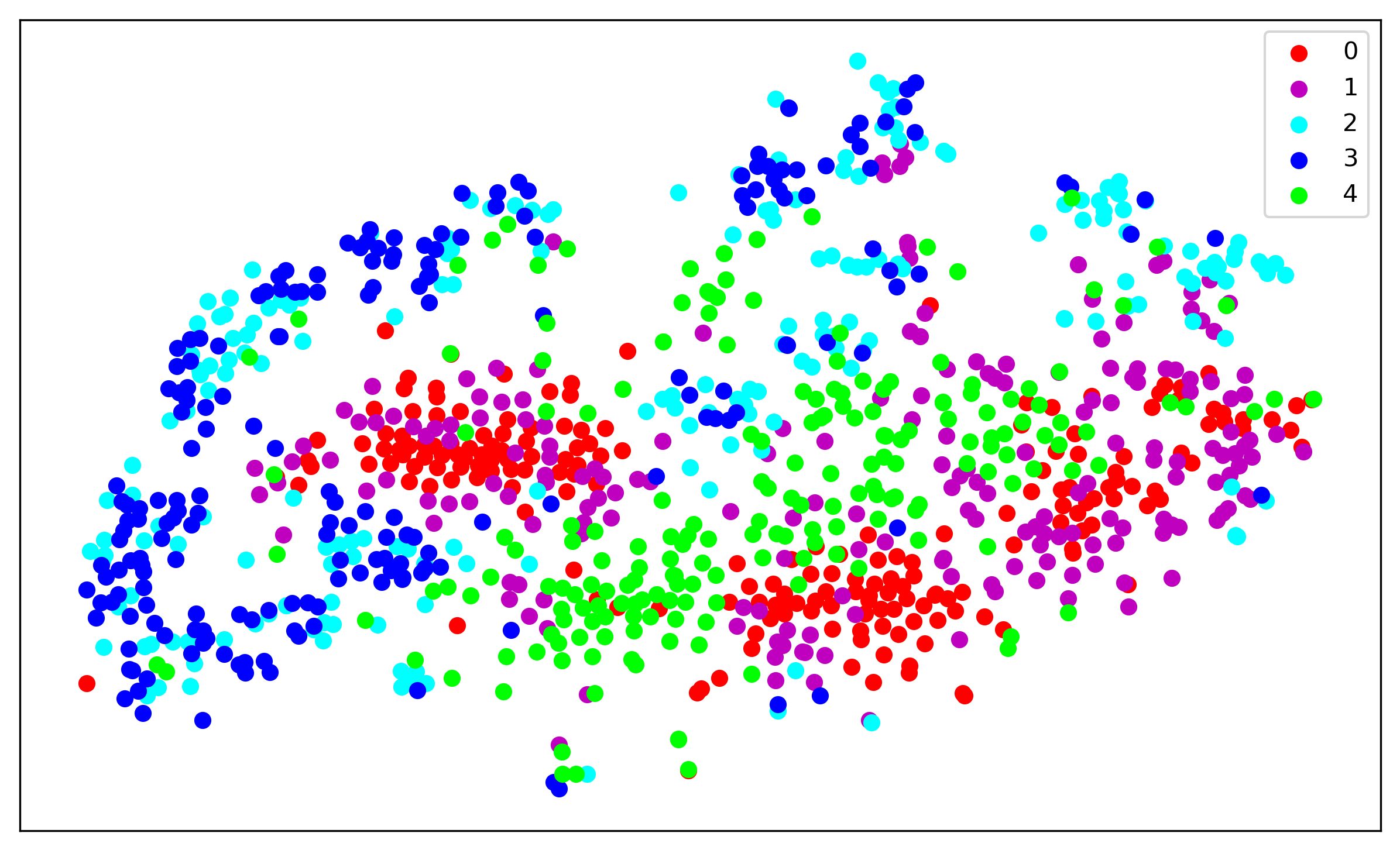}
\end{minipage}
}
\caption{Visualizations of raw samples. We color all points by their classes, and each point is a galaxy image.}
\label{fig:raw}
\end{figure*}

Figure \ref{fig:raw} presents visualizations of raw samples. We color all points by their classes, and each point is a galaxy image. Figure \ref{fig:train1000} shows the classes of galaxy have a poor separation and are quite tangled on training-1000 subset. Similar result on testing-1000 subset is shown in Figure \ref{fig:test1000} as well. It happens due to the absence of the use of supervised (labels) information before training. And both of the Figure \ref{fig:train1000} and Figure \ref{fig:test1000} indicate our dataset is complex and challenging.

\begin{figure*}
\centering
\subfloat[Map of the last fully connected layer of CNN 1 on the \newline training-1000 subset. (Accuracy: 94.9\%)]{
\label{fig:cnn1_train}
\begin{minipage}[t]{0.5\textwidth}
\centering
\includegraphics[scale=0.45]{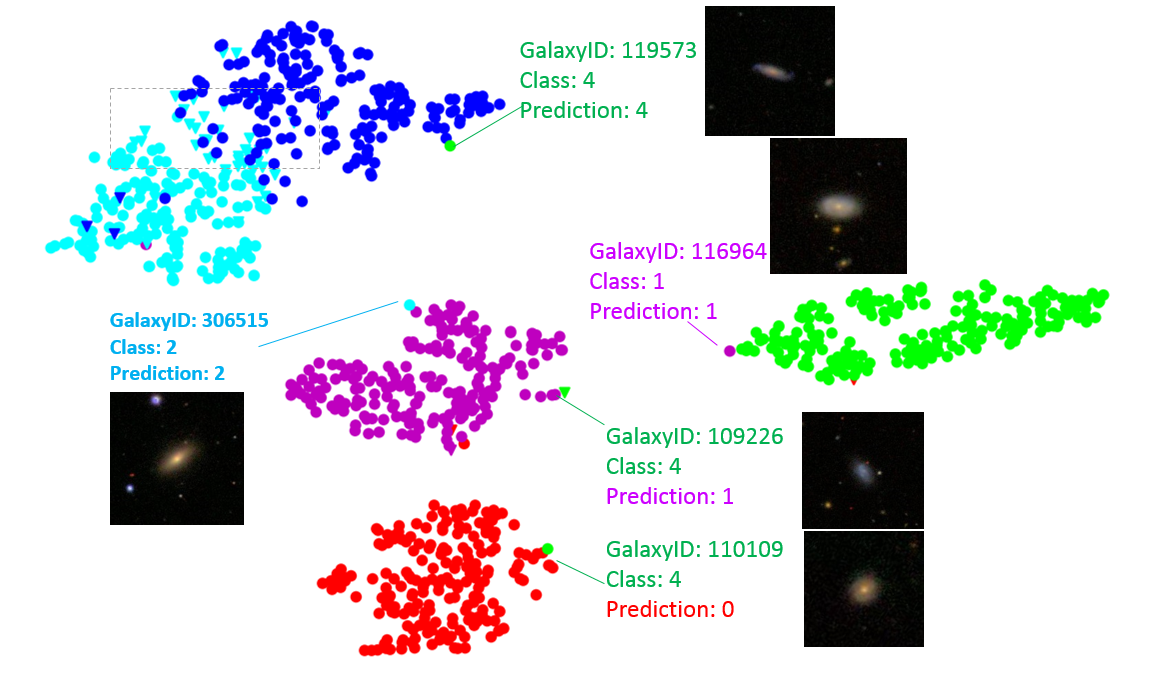}
\end{minipage}
}
\subfloat[Map of the last fully connected layer of CNN 1 on the testing-1000 subset. (Accuracy: 90.4\%)]{
\label{fig:/cnn1_test}
\begin{minipage}[t]{0.5\textwidth}
\centering
\includegraphics[scale=0.45]{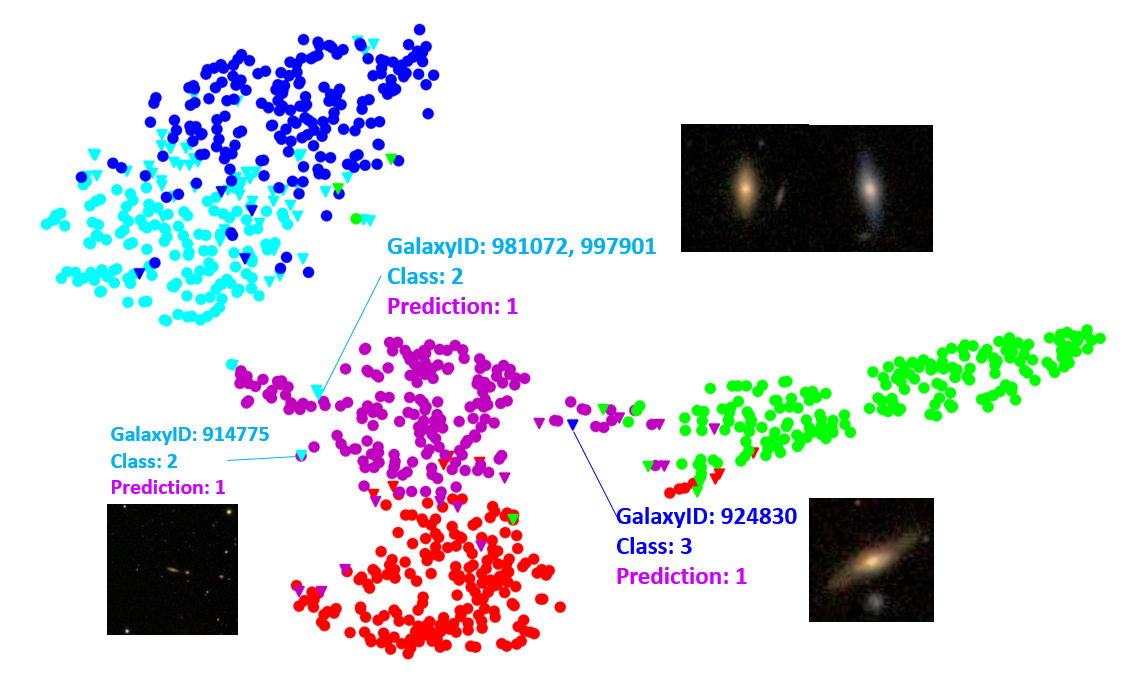}
\end{minipage}
}
\caption{Visualizations of activations of the last fully connected layer of CNN 1. We color all points by their classes, and mark misclassification by triangle glyphs. Each point is a galaxy image.}
\label{fig:cnn1}
\end{figure*}

Figure \ref{fig:cnn1} shows visualizations of activations of the last fully connected layer of CNN 1. We color all points by their classes, and mark misclassification by triangle glyphs. It can be seen that the visual separation among galaxy classes both on training-1000 subset and testing-1000 subset is improved significantly after training. As \citet{lecun2015deep} said, higher layers of representation amplify aspects of the input that are important for classification and discrimination. Figure \ref{fig:cnn1_train} and Figure \ref{fig:/cnn1_test} show that the last fully connected layer of CNN 1 has learned to transform the raw data into a nice representation and the classes are much more separated. From the visualizations, the images of each galaxy morphological class are presence of clusters, i.e. each class of galaxy images is grouped, the images with same morphological class are moved closer together. It can be explained that the images of same galaxy morphological class have similar underlying structure, so they tend to be grouped in scatter plots. From Figure \ref{fig:/cnn1_test}, we also can find the completely round and the in-between are closer, compared to other galaxy clusters. It is caused that the completely round and the in-between belong to a broad cluster, namely, smooth galaxy.

Then, inspecting the outliers is interesting. Consider the lime point placed in the bottom of blue cluster in Figure \ref{fig:cnn1_train}, which belongs to the spiral and is predicted to the spiral as well, but it has the similar structure with the edge-on. When inspected, the outlier sample (the spiral: GalaxyID 119573) looks very similar to the edge-on class. Many other outliers (the cigar-shaped: GalaxyID 306515 and the in-between: GalaxyID 116964) have the similar phenomenon in Figure \ref{fig:cnn1_train}. Another lime outlier near the red cluster is labelled as the spiral (GalaxyID: 110109), but is predicted herein to be the completely round. When inspected, it is really a completely round, i.e. the image numbered 110109 is incorrectly labelled. Figure \ref{fig:/cnn1_test} also shows some outliers like the cigar-shaped (GalaxyID: 981072, 997901) predicted to the in-between, the edge-on (GalaxyID: 924830) predicted to the in-between. After examination, we find the errors are so obvious that even human could recognize them incorrect merely by naked eyes without the aid of any machine.

\begin{figure*}
\centering
\subfloat[Map of the last fully connected layer of CNN 2 on the \newline training-1000 subset. (Accuracy: 87.1\%)]{
\label{fig:cnn2_train}
\begin{minipage}[t]{0.5\textwidth}
\centering
\includegraphics[scale=0.45]{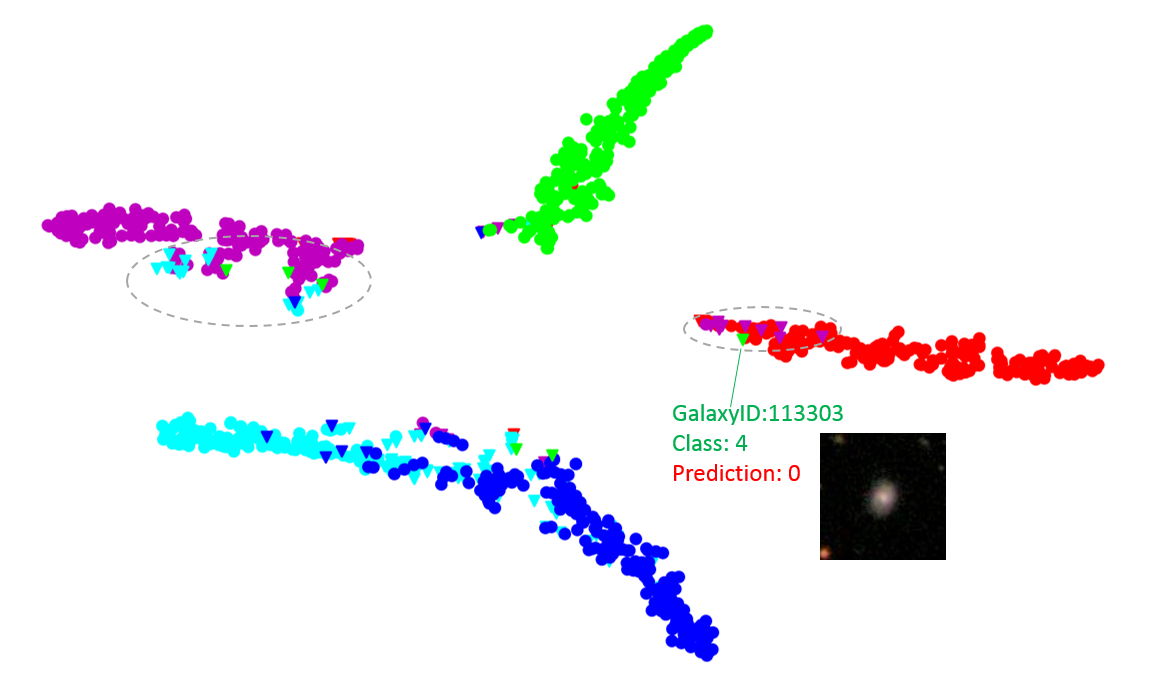}
\end{minipage}
}
\subfloat[Map of the last fully connected layer of CNN 2 on the testing-1000 subset. (Accuracy: 85.9\%)]{
\label{fig:cnn2_test}
\begin{minipage}[t]{0.5\textwidth}
\centering
\includegraphics[scale=0.45]{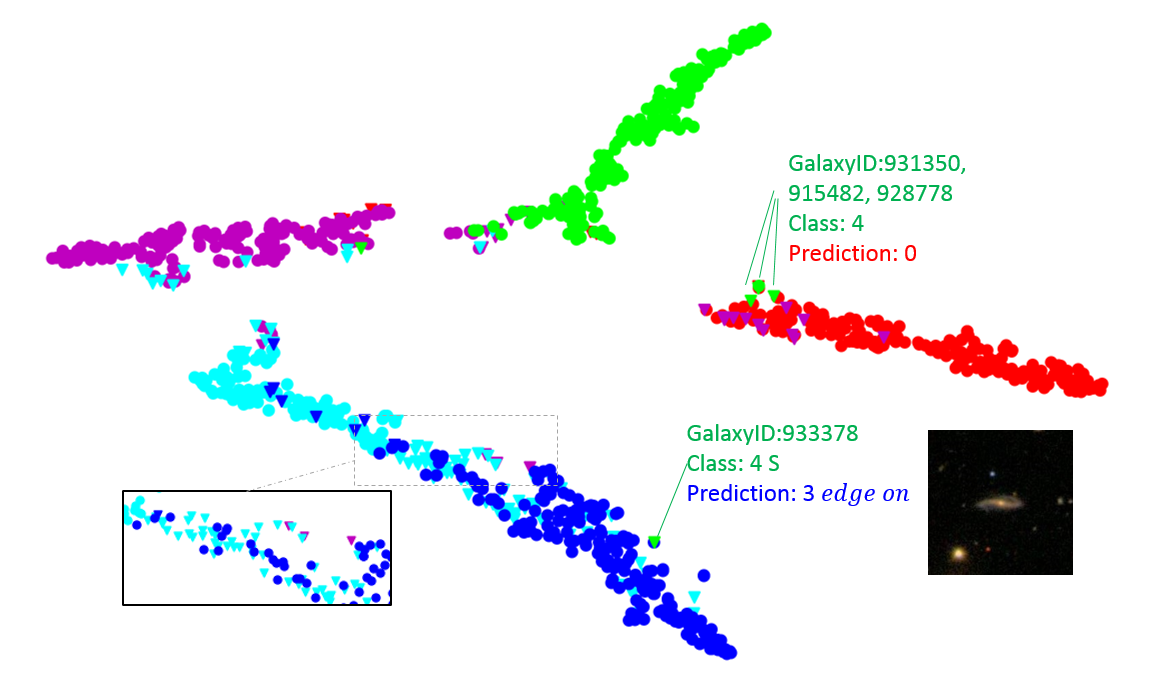}
\end{minipage}
}
\caption{Visualizations of activations of the last fully connected layer of CNN 2. We color all points by their classes, and mark misclassification by triangle glyphs. Each point is a galaxy image.}
\label{fig:cnn2}
\end{figure*}

\begin{figure*}
\centering
\subfloat[Map of the last average pooling layer of CNN 3 on the \newline training-1000 subset. (Accuracy: 91.9\%)]{
\label{fig:cnn3_train}
\begin{minipage}[t]{0.5\textwidth}
\centering
\includegraphics[scale=0.35]{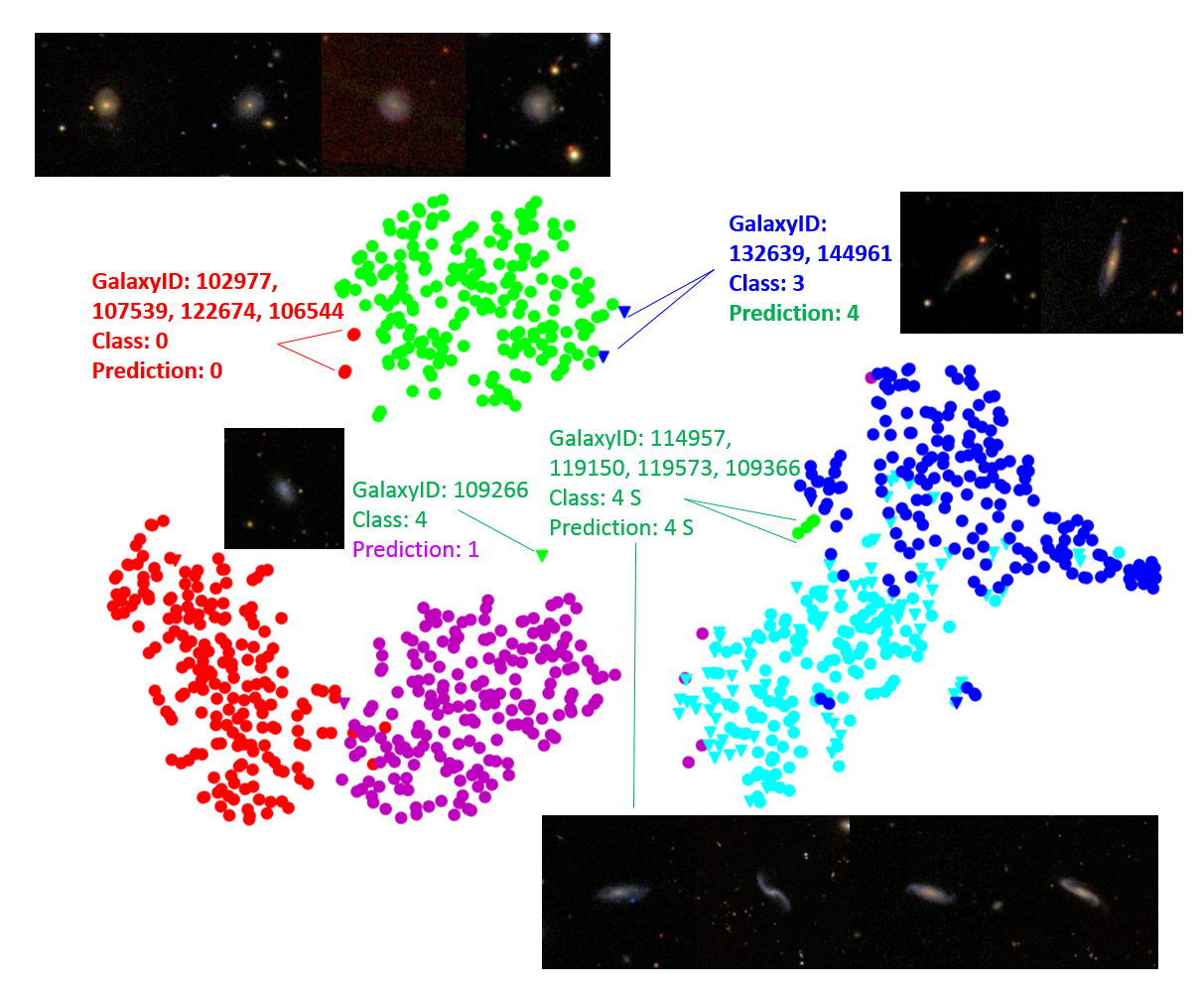}
\end{minipage}
}
\subfloat[Map of the last average pooling layer of CNN 3 on the testing-1000 subset. (Accuracy: 88.0\%)]{
\label{fig:cnn3_test}
\begin{minipage}[t]{0.5\textwidth}
\centering
\includegraphics[scale=0.4]{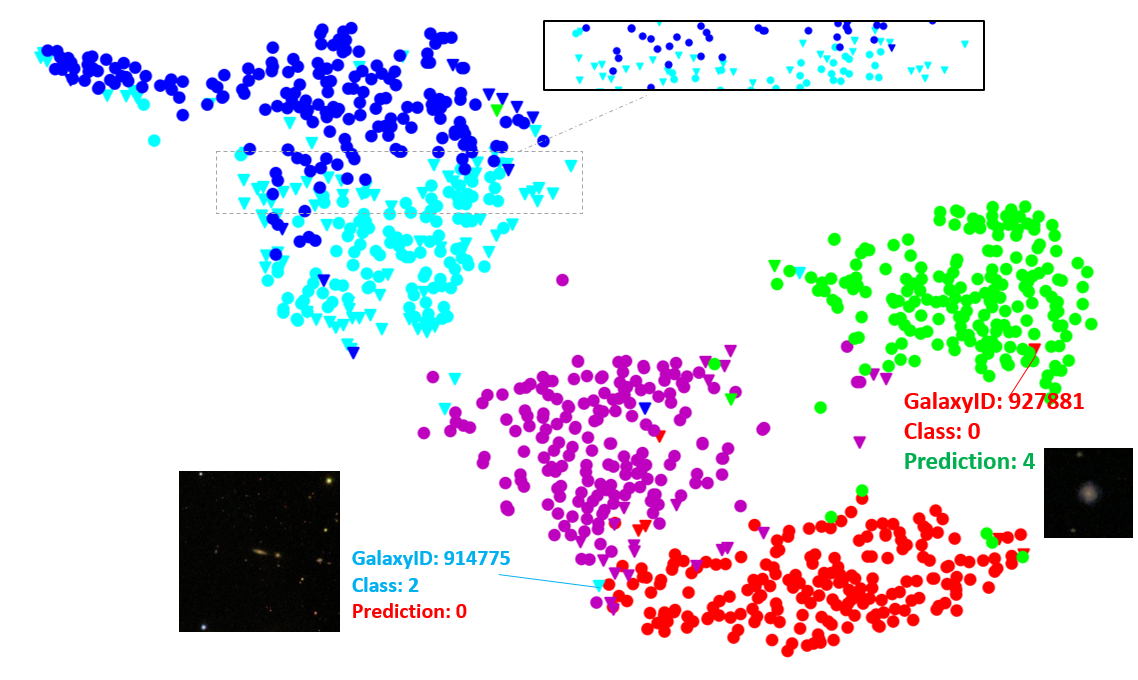}
\end{minipage}
}
\caption{Visualizations of activations of the last average pooling layer of CNN 3. We color all points by their classes, and mark misclassification by triangle glyphs. Each point is a galaxy image.}
\label{fig:cnn3}
\end{figure*}

Figure \ref{fig:cnn2} shows visualizations of activations of the last fully connected layer of CNN 2. Examining the Figure \ref{fig:cnn2_train} and Figure \ref{fig:cnn2_test}, we see that the separation between the galaxy classes is almost perfect and each class has a strip distribution. The two dotted ellipses show that 9 in-between are misclassified into the completely round and 16 cigar-shaped are misclassified into the in-between in Figure \ref{fig:cnn2_train}. This can be understandable that the completely round, the in-between and the cigar-shaped are all smooth galaxy. There is no exact limit to recognize them. We can say that they are misclassified, or they are wrongly labelled and classified correctly as well. The inset in Figure \ref{fig:cnn2_test} shows many cigar-shaped are misclassified into the edge-on.

Figure \ref{fig:cnn3} shows visualizations of activations of the last average pooling layer of CNN 3. Consider the red points (GalaxyID: 102977, 107539, 122674 and 106544) outlined in Figure \ref{fig:cnn3_train}, which correspond to the completely round and are predicted to the completely round as well, though are placed near the lime cluster corresponding to the spiral. After inspecting, there are three true completely round (GalaxyID: 102977, 107539 and 106544) but preserved the similar structure with the spiral, and one spiral galaxy (GalaxyID: 122674) labelled incorrectly. Another four spiral galaxies (GalaxyID: 114957, 119150, 119573 and 109366) are near the edge-on cluster and the cigar-shaped cluster. When inspected, they are all thin spiral galaxies and look very similar to the edge-on and the cigar-shaped. The outliers in Figure \ref{fig:cnn3_test} are similar to the outliers in Figure \ref{fig:/cnn1_test}.

There are 2 interesting galaxy images, numbered 109266 and 914775 (GalxyID). 109266 is labelled as the spiral in training-1000 subset, but is mistaken for the in-between both in Figure \ref{fig:cnn1_train} and Figure \ref{fig:cnn3_train}. After examining, we find it is a very faint spiral galaxy and very hard to recognize whether it is a spiral or others. 914775 is labelled as the cigar-shaped in testing-1000 subset, but is misclassified into the in-between in CNN 1 shown in Figure \ref{fig:/cnn1_test} and the completely round in CNN 3 shown in Figure \ref{fig:cnn3_test}, respectively. After inspecting, we find it is too small at the center of the images and really hard to recognize whether it is the cigar-shaped, the completely round or the in-between, which all belong to smooth galaxy.

The separation among the galaxy classes on training subset is better than on testing subset. It is reasonable because the training subset reaches a higher accuracy and learns a better representations.

Each small cluster like the completely round and the spiral is generally well grouped and almost perfect. And broad clusters like smooth galaxy including the completely round, the in-between and the cigar-shaped tend to be grouped closer together as well, as shown in Figure \ref{fig:cnn2_test}, Figure \ref{fig:cnn3_train} and Figure \ref{fig:cnn3_test}.

The visualizations reveal a fairly surprising phenomenon: the cigar-shaped and the edge-on are intertwined and overlapped, as shown in Figure \ref{fig:cnn1_train}, Figure \ref{fig:/cnn1_test}, Figure \ref{fig:cnn2_train}, Figure \ref{fig:cnn2_test}, Figure \ref{fig:cnn3_train} and Figure \ref{fig:cnn3_test}. This implies that the geometry of the cigar-shaped and the edge-on is similar. It is a common sense that the completely round, the in-between and the cigar-shaped are all smooth galaxies and similar, however, the results of visualization indicate that the cigar-shaped and the edge-on are much closed to together. This finding may help to further optimize the design of GZ2 decision tree.

\section{Conclusions}\label{sec:conclusion}

In this paper we train three CNNs to classify galaxies and learn automatically high-level abstract feature representations, use t-SNE techniques to visualize the high-dimensional feature representations in scatter plots, and explore the underlying structure of the data, and in some cases, the meaning of the data. Our experiments show that visualizations are of help to understand the global structure and the local structure of the galaxy, outliers, clusters with the learned feature representations extracted from CNNs. For example, the same morphological class galaxy images are clustered. A broad class, like smooth galaxy including the completely round, in-between and cigar-shaped, tends to group together and gets closer. It is an interesting phenomenon that the cigar-shaped and edge-on are intertwined. We find a completely round smooth galaxy is incorrectly labelled as spiral galaxy. This proves that the visualizations can help to find outliers in the dataset. It is hoped that this study could contribute to the exploring  and understanding of the galaxy image data itself with visualization, then provide valuable feedbacks to galaxy classification system, and facilitate the study of galaxy morphology. Our code is available for download at \url{https://github.com/Adaydl/GalaxyVisualization}.

In the future we plan to visualize galaxy datasets on  more fine-grained classes, e.g., 10 classes that would be more challenging. And we focus on more excellent deep learning algorithms to improve galaxy morphology classification performance and obtain higher quality galaxy feature representations.

\section*{Acknowledgements}

We would like to thank the galaxy challenge, Galaxy Zoo, SDSS and Kaggle platform for kindly sharing data. We acknowledge the financial support from the National Earth System Science Data Sharing Infrastructure (\url{http://spacescience.geodata.cn}). We are also supported by  CAS e-Science Funds (Grand XXH13503-04).




\bibliographystyle{mnras}
\bibliography{cit} 





\bsp	
\label{lastpage}
\end{document}